\documentclass[prl,twocolumn,showpacs,preprintnumbers,amsmath,amssymb]{revtex4}
\usepackage{graphicx}
\usepackage{dcolumn}
\usepackage{bm}
\usepackage{amsmath}
\usepackage{natbib}

\begin{document}

\title{\textbf X-ray Resonance in Crystal Cavities--Realization of Fabry-Perot Resonator for Hard X-rays}

\author{S.-L. Chang,$^{1,2,*}$ Yu. P. Stetsko,$^2$ M.-T. Tang,$^2$ Y.-R. Lee,$^1$ W.-H. Sun,$^1$
 M. Yabashi,$^3$ and T. Ishikawa,$^{3,4}$}
\affiliation{$^1$ Department of Physics, National Tsing Hua University (NTHU), Hsinchu, Taiwan,
 Republic of China 300.}
\affiliation{$^2$ National Synchrotron Radiation Research Center
(NSRRC), Hsinchu, Taiwan, Republic of China 300.}
\affiliation{$^3$ Spring-8/JASRI, Mikazuki, Hyogo 679-5198, Japan. }
\affiliation{$^4$ Spring-8/RIKEN, Mikazuki, Hyogo 679-5148, Japan.}

\date{\today}
\email[To whom correspondence should be addressed:]{slchang@phys.nthu.edu.tw}

\begin{abstract}
X-ray back diffraction from monolithic two silicon crystal plates
of 25--150 $\mu$m thick and a 40--150 $\mu$m gap using synchrotron
radiation of energy resolution $\Delta E=0.36$ meV at 14.4388 keV
shows clearly resonance fringes inside the energy gap and the
total-reflection range for the ($12$ $4$ $0$) reflection. This
cavity resonance results from the coherent interaction between the
X-ray wavefields generated by the two plates with a gap smaller
than the X-ray coherence length. This finding opens up new
opportunities for high-resolution and phase-contrast X-ray
studies, and may lead to new developments in X-ray optics.
\end{abstract}
\maketitle

Lasers of long wavelength, ranging from visible spectra to soft
X-rays, have had a great impact on the development of sciences and
led to diverse applications in physics, chemistry, biology,
material sciences, engineering, etc. In the case of shorter
wavelength hard X-rays, X-ray lasers, including free-electron
lasers, have long been anticipated to provide a coherent X-ray
source for probing structures of matter, periodic and
non-periodic, in sub-atomic scales, including the structures of
nano-particles, quantum dots, and single biological molecules. One
of the components for making lasers is the optical resonator, like
the Fabry-Perot resonators \cite{1}-\cite{3}. However, an X-ray
resonator for X-ray lasers has never been realized, although it
has long been proposed \cite{4}-\cite{6} and pursued
\cite{7}-\cite{14} from time to time for more than three decades.
In fact, the difficulty in realizing X-ray resonators with
observable resonance fringes arises mainly from the required
experimental conditions on coherence are not easily attainable,
aside from the many well documented theoretical studies of X-ray
cavity resonance \cite{3}-\cite{12} reported in the literature.
Recent experiments of Liss et al \cite{13} and Shvyd'ko et al
\cite{14} have observed storage of X-ray photons in a few tens
back-and-forth reflection cycles in two-plate crystals with
continuously decaying reflectivities in time-resolved experiments.
The latter have also showed the beating of two Pendell\"{o}sung
fringes \cite{15} on the tails of diffraction profiles in photon
energy scans, mainly due to crystal thickness effects. Although
these results are useful for X-ray optics and time-resolved
experiments, they did not show observable cavity resonance
fringes, mainly because the required coherence conditions were not
satisfied and the insufficient energy resolution used washed out
resonance fringes (see later discussion).

\begin{figure}[h]
\includegraphics[width=6cm]{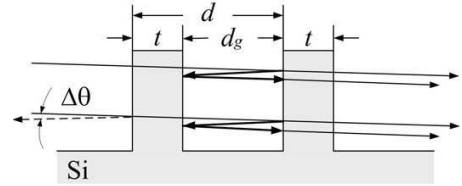}
\caption{Schematic of resonance interference in a Fabry-Perot type
two-plate cavity of silicon for an incident X-ray beam deviated
$\Delta\theta$ from the Bragg angle ($=90^\circ$.) for back
diffraction. Multiple reflections take place within the crystal
gap and generate forward-transmitted and back-reflected beams. The
coherent interaction among the transmitted and the reflected beams
inside the crystal plates and within the gap leads to cavity
resonance. \label{Fig. 1}}
\end{figure}

\begin{figure*}[t]
\includegraphics[width=12cm]{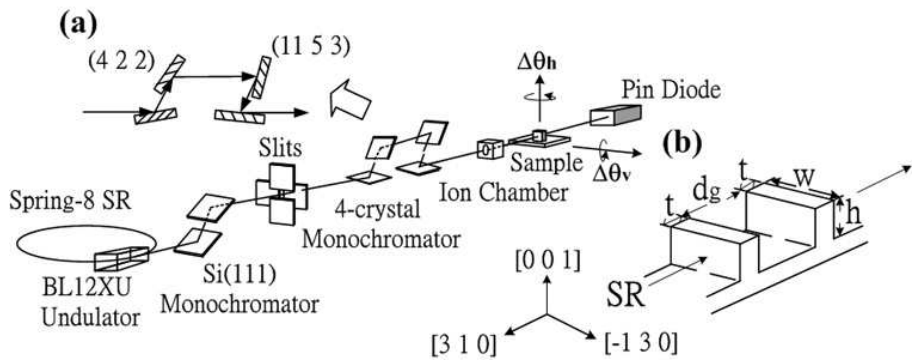}
\caption{Experimental set-up: (a) The side view of the
four-crystal high-resolution monochromator; (b) The
crystallographic orientation of the two-plate cavity.
$\Delta\theta_v$ and $\Delta\theta_h$ are the vertical and
horizontal tilting angles of the cavity crystal. The transmitted
diffracted and the back reflected beams are monitored by the pin
diode and ion chamber, respectively. \label{Fig. 2}}
\end{figure*}

An unambiguous way to demonstrate the cavity resonance is, under
appropriate coherent conditions, to show resonance interference
fringes inside the total reflection range and inside the energy
gap of a back diffraction \cite{7,11,12}. Moreover, images of
concentric interference rings are expected to be seen in X-ray
resonator as in an optical Fabry-Perot resonator \cite{1}. For an
optical Fabry-Perot resonator, the expected resonance interference
fringes in an energy scan should show well resolved fringes with a
separation $E_d=hc/(2d)$, the so-called free spectral range
\cite{1,2}, between fringes, where $h$, $c$, and $d$ are the
Planck constant, the speed of light, and the effective gap between
the two mirrors. If the spectral width of the fringe is $\Gamma$
and the energy resolution of the incident photon beam is $\Delta
E$, the required experimental criteria in energy scans for
observing well resolved resonance fringes are:(a) $\Delta E <
E_d$, (b) $\Delta E< \Gamma$, and (c)  $ \Gamma\leq E_d$. By
taking the reciprocal of (a), and the uncertainty principle,
$(\Delta t)(\Delta E)\cong \hbar$, the required criterion (d), for
time scans is $\Delta t(=\hbar / \Delta E))
>t_f/2\pi(=\hbar/E_d=2d/c/2\hbar)$, where $\Delta t$ is the coherent time
and $t_f$ the circulation time of photons between the two mirrors.
That means within the $\Delta t$ of the incident beam, photons in
the cavity can interact coherently. From criterion (d) and the
definition of longitudinal coherence length
$l_L=\langle\lambda^2/\Delta\lambda\rangle=\langle\lambda/(\Delta
E /E)\rangle$, it is easy to show that $l_L>2d$ (criterion (e)),
where $\langle\rangle$ means 'average'. For a normal distribution
the average equals to 1/2. From criteria (a)-(e) and the relations
between (a) and (d),(e), $\Delta E /E$ and $d$ are evidently the
key parameters, relating to the temporal coherence, which
determine whether resonance fringes can be observed. For an X-ray
cavity shown schematically in Fig. 1, the effective distance
$d=d_g+t$, where $d_g$ is the gap and $t$ the thickness of the
crystal plate.

Based on this consideration, we combine a small gap cavity of
40--150 $\mu$m prepared by the microelectronic technique and an
energy resolution of $\Delta E=0.36$ meV at $14.4388$ keV of X-ray
synchrotron radiation to fulfill the required coherence conditions
for cavity resonance and succeed for the first time in observing
resonance fringes inside the energy gap in energy scans and inside
the total-reflection range of angle scans near/at the ($12$ $4$
$0$) reflection position for silicon crystal cavities, thus
effectively realizing an X-ray resonator. Images of interference
rings are also observed, which again confirm the X-ray resonance
in crystal cavity for hard X-rays and near $\gamma$-rays. The
experimental results are reported in this Letter.

An X-ray Fabry-Perot resonator consists, in principle, of two
crystal plates as reflecting mirrors. Cavity resonance occurs when
an incident X-ray beam is reflected back and forth coherently
between the two plates (Fig.1), thus generating interference
fringes. Each reflection is a back diffraction from a set of
atomic planes with a Bragg angle very close to $90^\circ$. In this
study, several two- and multi-plate crystal cavities with a plate
thickness $t$ ranging from $25$ to 150 $\mu$m and a gap of 40--150
$\mu$m between them were prepared from a four-inch Si $(001)$
crystal wafer by using the microelectronic lithography process.
The patterned photoresist and mask were made on the wafer surface
according to the width and thickness of the crystal plate, and the
gap $d_g$ between adjacent crystal plates. The patterned surface
was then dry etched by reactive ions and cleaned afterwards. The
width $w$ and height $h$ of crystal plates are $800$ and 200
$\mu$m (fig. 2). Cavities with plate numbers up to $8$ were
manufactured. Sharp surfaces of each plate were seen from an
optical microscope.

\begin{figure}[h]
\includegraphics[width=8.5cm]{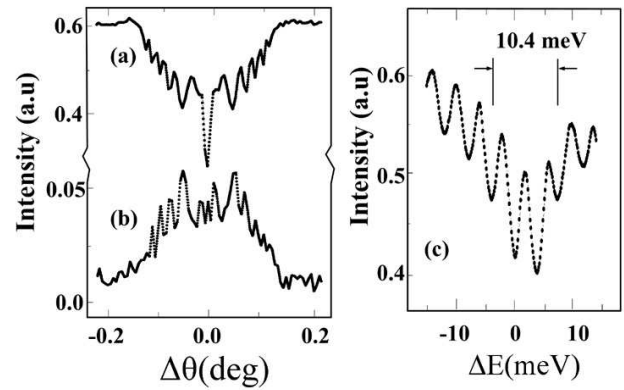}
\caption{$\Delta\theta_v$-scans of (a) the forward-transmitted
(000) beam and (b) the back-reflected ($12$ $4$ $0$) beam of the
two-plate cavity at $\Delta E = 9$ meV and $0.002^\circ$/step. (c)
The energy E scan. The rather noisy profile of Fig.3b is due to
the contribution of the incident beam and the poor resolution of
the ion chamber. The crystal cavity was kept in room temperature
with fluctuation less than $0.1^\circ$C. \label{Fig. 3}}
\end{figure}

The experiment was carried out at the Taiwan undulator beamline,
BL12XU, at the SPring-8 synchrotron radiation facility in Japan.
The storage ring was operating at $8$ GeV and $100$ mA. As shown
in Fig. 2, the incident radiation was monochromatized first by a
Si $(111)$ double-crystal and then by a four-crystal ultra-high
resolution monochromator for the energy $E=14.4388$ keV. The
four-crystal monochromator consisted of two pairs of $(422)$ and
($11$ $5$ $3$) asymmetric reflections in a ($+$$-$$-$$+$) geometry
similar to the reported arrangement \cite{16}. This monochromator
yields the energy resolution $\Delta E /E=$2.5x$10^{-8}$ at
$14.4388$ keV, i.e.,$\Delta E=0.36$ meV, such that the
longitudinal coherent length, $l_L=1717$ $\mu$m, is much longer
than the crystal gap (100$\mu$m). The criterion (e) for observing
cavity resonance is satisfied. Energy scans were performed by
tuning together the Bragg angles of the third and fourth crystals
with a minimum step of $0.005$ arcsecond, equivalent to 58.548
$\mu$eV in energy. The crystal cavity with the $[001]$ direction
along the vertical direction (Fig.2) was placed on a goniometric
head at the center of a Huber 8-circle diffractometer $1.5$ m away
from the last ($11$ $5$ $3$) crystal. The incident beam was in the
$[\bar{3}\bar{1} 0]$ direction normal to the large face of the
crystal plates. The beam size limited by slits and the
four-crystal monochromator was $0.05$ mm vertical and $0.2$ mm
horizontal at the crystal cavity. The rotations $\Delta\theta_h$
and $\Delta\theta_v$ of the crystal cavity around $[001]$ and
$[\bar{1}30]$ respectively (see Fig.2), with a minimum step of
$0.0005^\circ$, were performed with the diffractometer, and the
sample height was adjusted such that the incident beam hit only
the cavity.

\begin{figure}
\includegraphics[width=8.4cm]{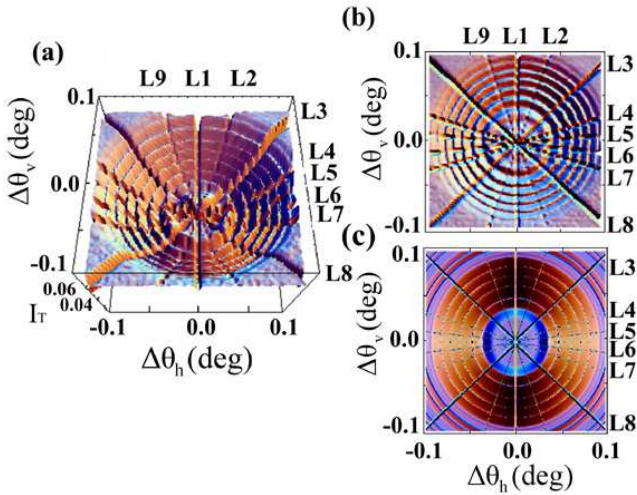}
\caption{Intensity distributions of resonance interference at
$\Delta E = 12$ meV:  (a) Angular ($\Delta\theta_h$,
$\Delta\theta_v$) distribution of the transmitted (000) intensity
$I_T$ of the two-plate crystal cavity in a linear scale; (b)
Two-dimensional contour map of Fig. 4a; (c) Calculated map of
(b)without angle integration. Multiple-beam interaction generates
additional radial intensity lines for 9 coplanar diffractions.
Lines L1 - L9 are related to the 9 coplanar diffractions C1 - C9,
where C1: (040), (4$\bar{4}$0), (480), (8$\bar{4}$0), (880) and
(12 0 0) reflections; C2: (6$\bar{4}$$\bar{2}$) and (682)
reflections; C3: (022) and (12 2 $\bar{2}$) reflections; C4:
(60$\bar{6}$) and (646) reflections; C5: (426) and (82$\bar{6}$)
reflections; C6: (42$\bar{6}$) and (826) reflections; C7: (606)
and (646) reflections; C8: (02$\bar{2}$) and (12 2 2) reflections;
C9: (6$\bar{4}$2) and (68$\bar{2}$) reflections. The (000) and (12
4 0) reflections are omitted in each of the coplanar diffractions
listed. (step width in $\Delta\theta_h$ and
$\Delta\theta_v:0.003^\circ$, counting time :1 sec.)\label{Fig.
4}}
\end{figure}

The ($12$ $4$ $0$) reflection was chosen as the back diffraction
for $E=14.4388$ keV. Both the forward-transmitted (000) and the
back-reflected ($12$ $4$ $0$) beams were monitored by a pin-diode
detector and an ion chamber, respectively (Fig. 2). Because of the
high energy X-ray used and the symmetry of the diamond structure,
$24$ reciprocal lattice points (r.l.p.s), including those of ($12$
$4$ $0$) and $(000)$ reflections, lie on the surface of the Ewald
sphere at this energy, thus generating 24 diffracted beams from
the center of the Ewald sphere towards these 24 r.l.p.s
\cite{17,18}. This 24-beam (simultaneous) diffraction consists of
the $9$ coplanar diffractions indicated in the legend of Fig.4,
each of which has the same zone axis. A two-plate cavity of 70
$\mu$m thickness with a 100 $\mu$m gap was aligned for the 24-beam
diffraction at $14.4388$ keV by adjusting the $\Delta\theta_h$ and
$\Delta\theta_v$. According to the dynamical theory of X-ray
diffraction \cite{15,18}, the angular $\Delta\theta_h$ and
$\Delta\theta_v$ scans of the forward transmitted beam $(000)$
show a broad dip profiles with an averaged flat bottom (see
Fig.3a) at the photon energies close to the exact $14.4388$ keV.
Figure 3b shows the $\Delta\theta_v$ scan of the back-reflected
($12$ $4$ $0$) beam. Similar profiles were observed for
$\Delta\theta_h$ scans (not shown). The region of broad width is
the total reflection region, which corresponds to the energy gap
in energy scans through the relation $\Delta E
/E=\Delta\theta^2/2$, where
$\Delta\theta=\sqrt{\Delta\theta_h^2+\Delta\theta_v^2}$ and
$\Delta E = E - 14.4388$ keV. The minimum intensity in the middle
of Fig.3a was due to the 24-beam diffraction. Distinct
interference fringes due to cavity resonance are clearly seen in
the transmitted (000) beam (Fig. 3a), and the back-reflected ($12$
$4$ $0$) beam (Fig. 3b). Figure 4a shows a 3-dimensional intensity
plot of a $\Delta\theta_h - \Delta\theta_v$ mesh scan of the
transmitted beam and Fig. 4b is the 2-dimensional projection. The
3-dimensional plot revealing the interference intensity
distribution is like a Roman amphitheater (Fig. 4a). The
2-dimensional fringes show concentric rings, i.e.,
$\Delta\theta=$constant, of alternating maxima and minima and the
straight lines are due to the coplanar diffractions mentioned. The
calculated 2-d image, Fig.4(c), agrees qualitatively with Fig.
4(b). At $\Delta\theta_v =0.03^\circ$ and $\Delta\theta_h
=-0.02^\circ$ slightly off the normal incidence position
($\Delta\theta_v =0^\circ$, $\Delta\theta_h =0^\circ$), an energy
scan was performed. The obtained profile near the energy gap of
$10.4$ meV is shown in Fig. 3c. The gap corresponds to the angular
width approximately equal to $3.6\sqrt{\chi}$  for the exact ($12$
$4$ $0$) back diffraction in angle scans for this crystal
thickness, $\chi \sim$ 0.4x$10^{-6}$ being the electric
susceptibility of ($12$ $4$ $0$). Again, interference fringes due
to cavity were observed. The fringe spacing, namely the $E_d$, is
about $3.60$ meV, in agreement with the calculated value $3.65$
meV from the relation $E_d=hc/2d$. This is also consistent with
the calculations based on the dynamical theory \cite{11,12,15,18}.
Since $\Delta E=0.36$ meV is smaller than $E_d=3.60 meV$,
criterion (a) is fulfilled. The effective resonance distance
$d=d_g+t$ can be considered as the distance between the two
electric fields, the X-ray wavefields, generated by each crystal
plate in one back-and-forth reflection period. Also,
$2(d_g+tn_x)\simeq2dn_x\simeq2d$ is the optical path length of
X-rays which gives a phase shift of $2\pi$, $n_x$ being the X-ray
index of refraction and $n_x=1-\delta$ with
$\delta=2.3$x$10^{-6}$. The spectral width $\Gamma =1.60$ meV of
fringes gives the finesse \cite{1,2,12} $F=E_d/\Gamma =2.3$, which
is less than the designed value $F=4.0$ due to the real
experimental situation, such as crystal surface roughness and
inclination, and small lattice distortion due to the sample
treatment and temperature effect. Various $\Delta\theta_h$ ,
$\Delta\theta_v$, and energy scans were carried out for other
2-plate and 8-plate cavities and similar interference fringes were
observed. Here $\Delta E=0.36$ meV $<\Gamma =1.60$ meV, and $
\Gamma=1.60$ meV $\leq E_d=3.60$ meV, so criteria (b) and (c) are
also fulfilled.

The time $t_f$ for X-ray photons travelling between the two
crystal plates of 100 $\mu$m gap is $0.67$ ps, much shorter than
the time $\Delta t (=1.8$ps$)$. Namely, $\Delta t
>t_f/2\pi (=0.1$ ps$)$. Therefore, criterion (d) is satisfied. Moreover, as
already stated, $l_L>2d$, the longitudinal (temporal) coherence is
maintained. The transverse coherence determined by the photon
emittances is also retained for X-rays very close to normal
incidence. Hence both the spatial and temporal coherence of the
X-ray beams guarantee the realization of X-ray cavity resonance.
In contrast, the experimental conditions of the previous work
\cite{14}: $\Delta E(=2 $ meV $)$$
> E_d (=12.4 \mu$ eV$)$, $\Delta E(=2$ meV$) > \Gamma (= 0.64 \mu$ eV$)$,
$\Delta t (=0.33$ ps$) < t_f/2\pi (=53 $ps$)$, and $l_L(=0.31$
mm$)<2d(=100$ mm$)$, do not fulfill the required criteria (a),
(b), (d), and (e). Similarly, these required conditions were also
not satisfied by the work of \cite{13}. Under these unfavorable
circumstances, the expected resonance fringes spaced in 12.4 $\mu
eV$ intervals would be certainly smeared out by the energy
resolution of 2 meV\cite{14}.

For different photon energies, different back diffraction together
with different multiple reflections needs to be employed. The
presence of multiple diffraction is not a problem. Slight angular
deviation of the crystal cavity from the multiple-beam positions
can always be achieved without losing the resonance condition.
Since the fixed phase relation between the forward transmitted and
the back-reflected beams and the narrow energy and angular widths
of X-rays from the cavity, this crystal cavity with a better
finesse (i.e., thicker crystal plates with smaller surface
roughness) can be used for phase-contrast and high-resolution
X-ray optics, such as high-resolution monochromator using the back
reflection and narrow-band filter with the transmission
\cite{7,11,13,14}. All of these can be applied for high-resolution
X-ray scattering, spectroscopy, and phase-contrast microscopy in
many physical, chemical, and biological studies, such as
investigation of dynamics of solids, liquids, and bio-molecules,
precise measurements of wavelength and lattice constant, etc
\cite{7,11,13,14,17,19}. Furthermore, improved crystal cavities of
the present type with better finesse might be useful for the
development toward hard X-ray (or $\gamma$-ray) lasers, if
suitable lasing materials \cite{20} could be developed.

In conclusion, we have successfully observed cavity resonance
fringes in silicon crystal cavities and realized for the first
time Fabry-Perot resonators for hard X-rays (near $\gamma$-rays).
The required conditions for cavity resonance in normal incidence
geometry are the criteria (a)-(e), with which the longitudinal
coherence is retained for X-ray photons in the cavity.  The
quasi-coherent and highly monochromatic X-rays generated from the
resonator provide new opportunities for X-ray optics,
spectroscopy, and microscopy.

\begin{acknowledgments}
We thank B.-Y. Shew of the NSRRC and Y.-H. Lin of the Precision Instrumentation
Development Center for the preparation of the crystal resonators, K. Tamasaku and
D. Miwa of Spring-8/RIKEN and Y.-C. Cai, C.-C. Chen of NSRRC for technical supports,
J.-T. Shy of NTHU for discussion, and the Ministry of Education and National Science
Council of Taiwan, R.O.C. for financial supports.
\end{acknowledgments}

\bibliography{basename of .bib file}

\end{document}